# The Effects of Space Weather on Flight Delays

Y. Wang[1], X. H. Xu[1], F. S. Wei[1], X. S. Feng[1], M. H. Bo[2], H.W. Tang[2], D. S. Wang[2], L. Bian[2], B.Y. Wang[1], W. Y. Zhang[1], Y. S. Huang[1], Z. Li[3], J. P. Guo[4], P. B. Zuo[1], C. W. Jiang[1], X.J. Xu[5], Z. L. Zhou[5] and P. Zou[1]

1 Institute of Space Science and Applied Technology, Harbin Institute of Technology, Shenzhen, China
2 Travelsky Mobile Technology Limited, Beijing, China
3 Nanjing University of Information Science and Technology, Nanjing, China
4 Beijing Normal University, Beijing, China
5 State Key Laboratory of Lunar and Planetary Sciences, Macau University of Science and Technology, Macao, China

**Abstract**

Although the sun is really far away from us, some solar activities could still influence the performance and reliability of space-borne and ground-based technological systems on Earth. Those time-varying conditions in space caused by the sun are also called space weather, as the atmospheric conditions that can affect weather on the ground. It is known that aviation activities can be affected during space weather events, but the exact effects of space weather on aviation are still unclear. Especially how the flight delays, the top topic concerned by most people, will be affected by space weather has never been thoroughly researched. By analyzing huge amount of flight data (~$5 \times 10^6$ records), for the first time, we demonstrate that space weather events could have systematically modulating effects on flight delays. The average arrival delay time and 30-minute delay rate during space weather events are significantly increased by 81.34% and 21.45% respectively compared to those during quiet periods. The evident negative correlation between the yearly flight regularity rate and the yearly mean total sunspot number during 22 years also confirms such delay effects. Further studies indicate that the interference in communication and navigation caused by geomagnetic field fluctuations and ionospheric disturbances associated with the space weather events will increase the flight delay time and delay rate. These results expand the traditional field of space weather research and could also provide us with brand new views for improving the flight delay predications.

## Introduction

Dramatic variations in electromagnetic fields and plasma conditions above the Earth during space weather events (SWEs) are able to affect numerous aspects of our human society[1, 2]. Solar Flares (SFs), Coronal Mass Ejections (CMEs), and Solar Energetic Particles (SEPs) are typical SWEs[3, 4]. SFs bring Earth with violently increased electromagnetic radiation across the entire electromagnetic spectrum, from radio waves to gamma rays. These enhanced radiations will increase the ionization of the atmosphere, disturb the ionosphere and may cause radio blackout, especially for the aircraft High Frequency (HF) communication[5-7]. CMEs behave as dense plasma clouds that hit and compress the magnetosphere to disturb the near-earth space. Geomagnetic storms can also be triggered and these storms associated global geomagnetic field fluctuations and ionospheric disturbances will interfere with communication, navigation, and electric power transmission[8-10]. SEPs are very high-energy particles, some of which can even penetrate into the troposphere. These particles can strike aircraft electronics to cause single-event error that damage the avionics systems and reduce the safety margin of aircraft systems[11, 12]. In addition, the SEPs associated ionizing radiation could also make the crews and passengers exposed to excessive radiation environment especially near the polar region[13-16]. Although the occurrence of SWEs is highly related to the 11-year solar cycle, SWEs may occur every day. When SWEs reach the Earth, their influence could usually last from 1-3 days[1-4].

Analyzing the interrelationships between space weather and aviation is a new and developing research topic[13, 17, 18]. In the past few years, many scholars and international communities have paid more and more attention to the impact of space weather on the aviation industry[17, 19]. It is reported that space weather could affect terrestrial weather activities such as thunderstorms and lightning, which would have direct effects on flight safety[20]. Some analysis even revealed that SWEs could contribute to aviation accidents[21]. It is also noted that some flights have to change their schedules, routes or lower their cruising altitude to avoid radiation hazards during SEPs[15, 16]. In 2011, the International Air Transport Association (IATA) has realized the importance of acquiring space weather information, while early in 2002, the International Civil Aviation Organization (ICAO) has begun to evaluate the necessity to provide space weather information for international air navigation during SFs, and now ICAO provides real-time and worldwide space weather updates for commercial and general aviation to help ensure flight safety.

Nowadays, the growing concerns about space weather in aviation industry are mainly due to the safety issues, because safety is indeed the backbone of the aviation[17, 19, 22]. However, the real impacts of space weather on aviation could go far beyond the safety issues[3, 4]. We still don't know what aspects of aviation would be affected by space weather and how space weather would influence certain aspects of aviation. Particularly, how flight delays are influenced by space weather has not been thoroughly researched. How to quantify the space weather impacts on flight delays? What is the internal relationship between them? In this letter, we comprehensively investigate the effects of space weather on flight delays for the first time. It is revealed that the flight delay time and delay rate will be systematically modulated by SWEs. The clear correlations between flight delays and magnetospheric-ionospheric disturbances could bring us new ideas to help prevent or cope with flight delays.

**Method and Data**

It should be noted that flight delays have their own distinctive characteristics. For example, flight delays will be severely affected by some contingencies, such as air traffic congestion, inclement weather, or other security issues. In addition, the delay rate has its interior periodicity. It behaves quite differently from morning to midnight during one day, and it also changes during different weekdays and seasons. Therefore, in order to derive the 'real' effects of SWEs on flight delays, the influences of other factors should be carefully considered.

To start with, we divide the flight data into two comparative groups: SWEs affected flights and quiet time flights. The space weather affected periods are defined as the time from the beginning of a SWE (regardless of SF, SEP or CME) reaching the Earth till the next 24 hours, and any flight on the voyage in this period is considered to be affected by SWEs. The quiet time periods (QTPs) refer to the complete days (from 00:00 to 24:00, local time) when there are no SWEs, and any flight whose real takeoff and landing time are all in QTPs is defined as quiet time flights. The choice of 24-hour duration can avoid the daily periodicity of the flight delays, and the random distributions of SWEs in hours, days, weekdays and months in Fig.1 also suggest that the following deduced results would not be affected by the internal periodicities of the flight delays. Moreover, 5-year's huge amounts of flight data ($\sim 5 \times 10^6$ records), far beyond similar studies[23], are also used which could smear out the contributions of various contingencies as much as possible.

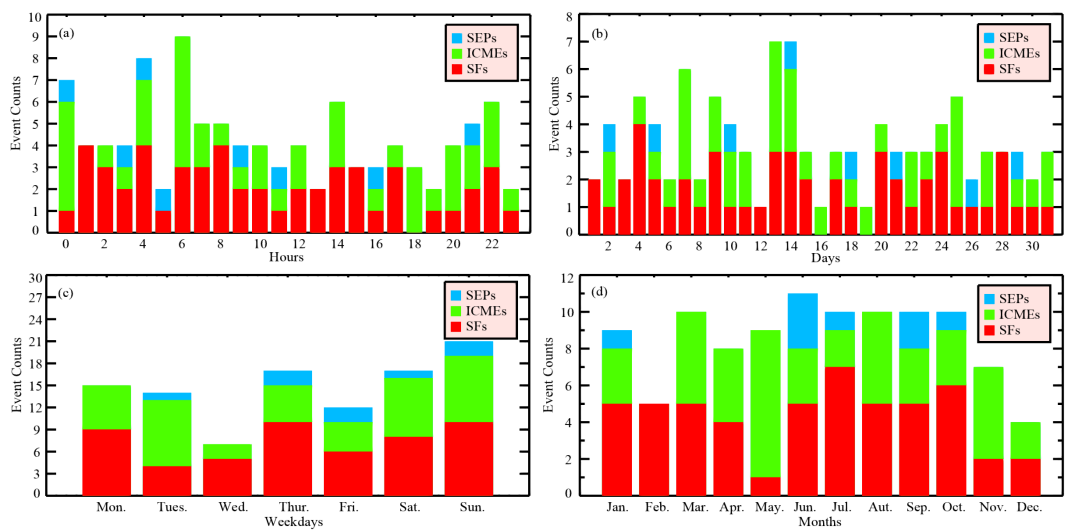

Fig.1. The distributions of SWEs in hours, days, weekdays and months from 2015 to 2019.

The individual flight data used in this paper is provided by the Travelsky Mobile Technology Limited, an affiliated company of Civil Aviation Administration of China (CAAC), who has all the commercial flight takeoff and landing records of Chinese airlines. Due to historical reasons, intact flight records are only available from 2015 in China. To avoid the impacts of Covid-19 on aviation, the investigated flight data is set from January 1st, 2015 to December 30th, 2019. All valid commercial flight data (totally 5,333,353 flight records) in the five largest hub airports in China, namely,

Guangzhou Baiyun International Airport (CAN), Shanghai Pudong International Airport (PVG), Beijing Capital International Airport (PEK), Shanghai Hongqiao International Airport (SHA) and Shenzhen Baoan International Airport (SZX), are selected in our analysis. Meanwhile, the national flight regularity rate from 1998 to 2019 is obtained from the CAAC Civil Aviation Development Statistics Bulletin.

The SWEs discussed here refer in particular to SFs, CMEs and SEPs. Only M-class and X-class SFs are selected and the events list are mainly obtained from the National Oceanic and Atmospheric Administration (NOAA) website, while the soft X-ray data from GOES satellites are also used to help identify the SFs. The CMEs are selected through the ICMEs list compiled by Richardson and Cane[24], and the SEPs are directly obtained from the NOAA Space Weather Prediction Center. Finally, 103 SWEs from January 1st, 2015 to December 30th, 2019 are selected to match to flight records and the complete SWEs list is provided in the Appendix. The disturbance storm time (Dst) index data are obtained from the International Service of Geomagnetic Indices. The total electron content (TEC) data are derived from the Madrigal database. The critical frequency of ionospheric F2 layer (foF2) are got from the Meridian Project Data Center. To make the letter compact, the related data processing methods will be shown in the Appendix.

## Results

The detailed distributions of flight arrival delay time (difference between the actual and schedule gate in time) are illustrated in Fig. 2. It can be seen that the distributions of flight delays are obviously altered by the SWEs. Although their distributions display the similar shape, the distributions of delay time affected by SWEs tend to shift to the right as a whole, which means that the delay time would be larger during SWEs than those in the QTPs. Specifically, the two delay distributions intersect near -10-0 minutes, and the probability density of positive delays during SWEs is always larger.

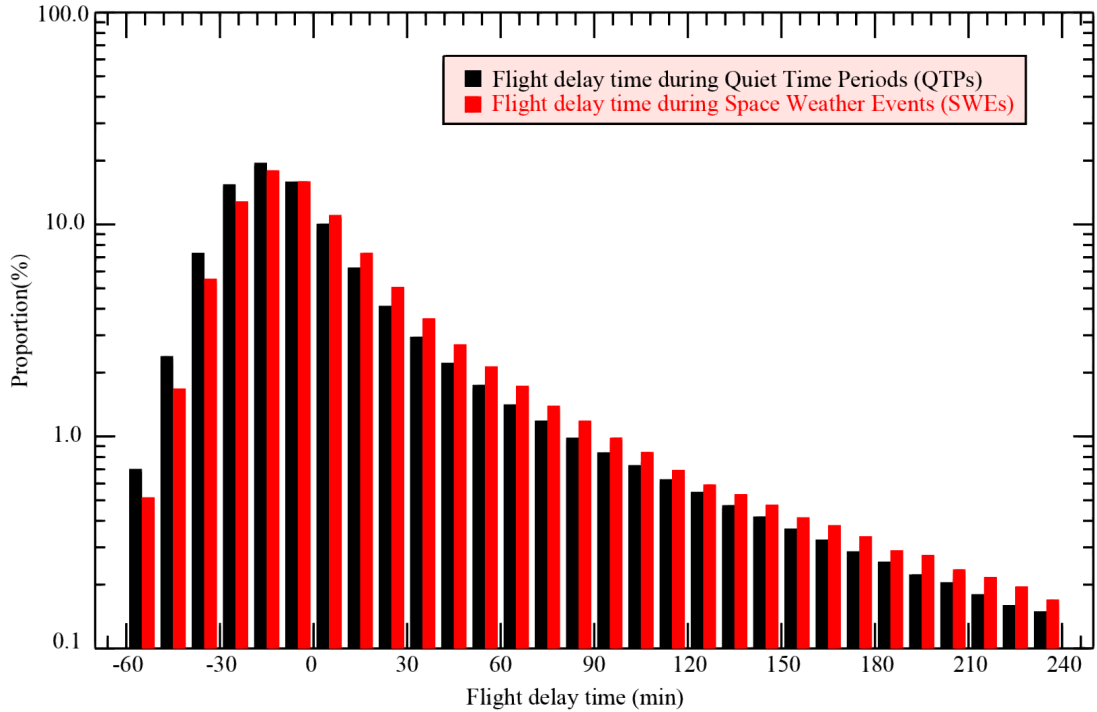

Fig.2 Probability distributions of flight arrival delays during SWEs (red) and quiet time periods (blue) averaged over 5 airports. The results of SWEs are calculated by weighted average of SFs, CMEs and SEPs.

To clearly quantify the effects of space weather on flight delays, the arrival delays during SWEs together with their increments relative to those during QTPs are listed in Table 1. It is found that the average arrival delay time during SWEs is substantially increased by 81.34% compared with those during QTPs (from 9.11min to 16.52 min). The 30-minute arrival delay rate increases from 17.76% to 21.57%, while the long-term (≥240-minute) delay rate increases by 47.59% during SWEs. It is noteworthy that SWEs have more remarkable influences on the delay time than the delay rate. Moreover, both the largest delay time and the highest delay rate seem to suggest that the SEPs have the most obvious effects on flight delays among the three types of SWEs.

Table 1 Arrival delay time, ≥30-minute delay rate and ≥240-minute delay rate of flights during SFs, CMEs, SEPs, and all SWEs together with their (relative) increments relative to those during quiet time periods averaged over 5 airports.

|  | Average delay time | | | ≥30 min delay rate | | | ≥240 min delay rate | | |
| --- | --- | --- | --- | --- | --- | --- | --- | --- | --- |
|  | Delay time (min) | Increment (min) | Relative Increment (%) | Delay rate (%) | Increment (%) | Relative Increment (%) | Delay rate (%) | Increment (%) | Relative Increment (%) |
| QTPs | 9.11 | - | - | 17.76 | - | - | 1.45 | - | - |
| SFs | 17.24 | 8.13 | 89.24 | 22.15 | 4.39 | 24.72 | 1.95 | 0.5 | 34.48 |
| CMEs | 13.91 | 4.8 | 52.69 | 19.84 | 2.08 | 11.71 | 2.10 | 0.65 | 44.83 |
| SEPs | 28.39 | 19.28 | 211.64 | 27.45 | 9.69 | 54.56 | 3.60 | 2.15 | 148.28 |
| SWEs* | 16.52 | 7.41 | 81.34 | 21.57 | 3.81 | 21.45 | 2.14 | 0.69 | 47.59 |

*Note: the results of SWEs are calculated by weighted average of SFs, CMEs and SEPs.

For the first time, the above results indicate that SWEs could have nonnegligible systematical effects on flight delays. However, the internal relationship between SWEs and flight delays has never been studied thoroughly. As stated in the introduction section, previous researchers have suggested that SWEs could have many negative effects on high-tech systems on Earth[7, 11-13, 16, 19, 22]. Among these various harmful effects, and taking the reality of the aviation industry into consideration, we propose that the impacts of SWEs on communication and navigation should be given the most attention.

Communication and navigation form the key functions in modern air traffic management, and they are cornerstones that ensure the safety and efficiency in air traffic. Either the degradation or interruption of communication or navigation, whether on the air routes or near the airports are common reasons for flight delays. While malfunctions of communication and navigation system could be directly attributed to the geomagnetic field fluctuations and ionospheric disturbances driven by SWEs[5, 6, 9, 25].

Analyzing the Dst data, foF2 data and TEC data are common research methods for quantifying the disturbances in geomagnetic field and ionosphere[8, 26, 27]. The Dst index could represent the change of the horizontal component of the Earth's magnetic field, and the sudden change in Dst is a characteristic of geomagnetic storm[28]. During geomagnetic storms, the violent changes in magnetic fields could induce destructive currents that might disrupt the electric-power systems on the ground. Moreover, the geomagnetic storm could trigger ionospheric storm that greatly disturb the ionospheric environment[3, 4]. The foF2 is very important in HF communications since the HF signal needs ionosphere for reflection over long distances. The radio signal would be absorbed excessively and its propagation path could become unexpectedly when the foF2 is disturbed, so the communication quality would be degraded or even interrupted during ionospheric disturbances. To evaluate such effects, the deviation of the monthly median of foF2 (ΔfoF2) is introduced to quantify the normal fluctuations of the ionosphere relative to its undisturbed status. In that case, ΔfoF2 could also be used to evaluate the communication quality[3, 4, 26]. The Rate of TEC index (ROTI), defined as the standard deviation of the rate of change of the TEC, is another useful indicator to describe the temporal ionospheric irregularities. Ionospheric irregularities are usually small-scale disturbances in the ionosphere that could lead to significant interferences of many satellite-related systems by rapidly modifying the amplitude and phase of a radio signal (ionospheric scintillation). Consequently, these disturbances in the ionosphere would also affect communication and navigation systems, such as the Global Navigation Satellite System (GNSS) [3, 4, 29]. Therefore, to investigate the internal relationships between SWEs and flight delays, here we would consider the rate of change in Dst (dDst), ΔfoF2 and ROTI as three most important indicators and try to reveal the impacts of geomagnetic field fluctuations and ionospheric disturbances on flight delays.

The time resolution of the Dst index is 1 hour, so the dDst in a certain time (ti) is the absolute difference of the Dst in the adjacent two hours, $dDst_{ti}=|Dst_{ti} - Dst_{ti-1}|$. ΔfoF2 is a dimensionless quantity, $\Delta foF2_{ti}=|foF2_{ti} - foF2_{med}|/ foF2_{med}$, where $foF2_{med}$ is the moving median of the nearby 28-day data. $ROTI_{ti} = \sqrt{< ROT_{ti}^2 > - < ROT_{ti} >^2}$,

where the < > denotes moving average during 1 hour and $ROT_{t_i}=(TEC_{t_i} - TEC_{t_{i-1}})/(t_i - t_{i-1})$.

The geomagnetic field fluctuations and ionospheric disturbances indicated by dDst, ΔfoF2 and ROTI, together with their relations to flight delays are shown in Fig.3. It is found in Fig.3(a) that when geomagnetic field is in a relative stable period (dDst<~10 nT/hour), both the delay time and delay rate increase slightly with dDst. As the fluctuations of geomagnetic field graduate to a more intense stage (especially dDst >~20 nT/hour), the flight delays also become obviously larger. Similar phenomena could be found in Fig.3(b) and (c) that both the delay time and delay rate tend to show roughly positive relationships with ionospheric disturbances. In particular, no obvious delay increases are seen when ΔfoF2<15%, while good monotonically increasing linear relationship between ΔfoF2 and flight delays are found when ΔfoF2>15%. As to the ROTI, it is found that the delay time and delay rate reveal sharp-gentle-sharp increases with ROTI, and the most prominent turning point is around 0.2 TECU/min. Regardless of dDst, ΔfoF2 and ROTI, all the similar behaviors in flight delays indicate that the flight delays would not be substantially influenced by the magnetospheric-ionospheric disturbances when the disturbances is relatively small. However, when the disturbance exceeds a certain threshold (e.g., dDst >~20 nT/hour, ΔfoF2>15% and ROTI>0.2 TECU/min), the flight delays display conspicuous positive monotonic relationships with the degree of geomagnetic field fluctuations and ionospheric disturbances. Fig.3(d), (e) and (f) also shown that the probability density of dDst, ΔfoF2 and ROTI during SWEs are always higher on the right-hand side, which implies that the degree of geomagnetic field fluctuations and ionospheric disturbances is obviously larger during SWEs than those during QTPs. These results are consistent with previous studies[8, 9], and all these analyses indicated that the SWEs will make flight delays more serious.

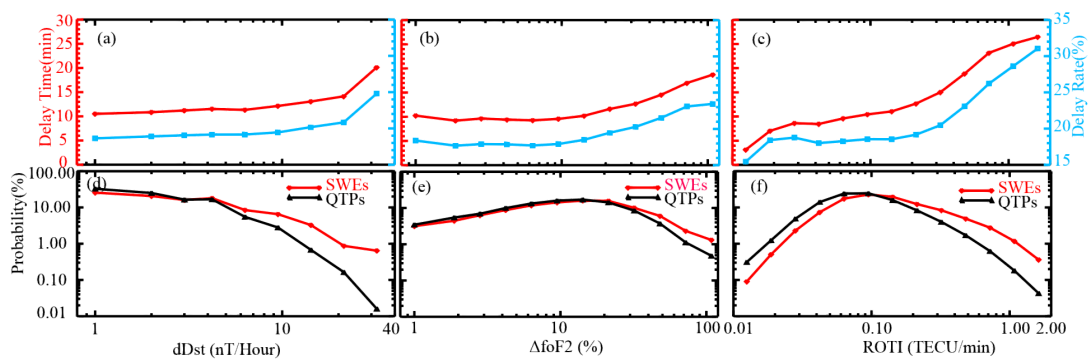

Fig.3 The distributions of flight delay time (red) and 30-minute delay rate (blue) as a function of dDst (a), ΔfoF2 (b) and ROTI (c) calculated from all flight records. The probability distributions of dDst, ΔfoF2 and ROTI during SWEs (red) and QTPs (blue) are also shown in panel (d), (e), (f).

## Discussions and Conclusions

Eliminating various interference factors is quite important for deriving the statistically valid effects of SWEs on flight delays. Actually, although presented statistical work has avoid the daily periodicity of flight delays, the impacts of space weather on flight delays are still underestimated. First of all, the real impacts of SWEs on our Earth can persist far beyond 24 hours. For example, the SWEs triggered geomagnetic storm can even last for 3-5 days, so a part of SWEs affected flights will be classified into QTPs. Such scenario can also be revealed in Fig.3 (d), (e) and (f). One can find that there still exists large value of dDst, ΔfoF2 and ROTI during QTPs, although their probability density is relatively very low. In addition, both geomagnetic field fluctuations and ionospheric disturbances need time to respond to SWEs[30]. However, some SWEs affected flights in current definitions, especially those at the beginning of a SWE, may not really be affected by SWEs. Therefore, in real situations, the SWEs should have more significant effects on flight delays, and the delay difference between SWEs and QTPs will be more prominent.

Certainly, using 48-hour or 72-hour duration to define SWEs can also avoid the flight's daily periodicity. Nevertheless, selection of a longer duration would lead to overlapping events since SFs, SEPs and CMEs often arrive on Earth successively. In presented analysis, the selection of 24-hour duration of SWEs is an acceptable choice after deliberations that could both guarantee the independence of each event and achieve a reasonable result. While five out of eight investigated SEPs cannot yet be isolated since they arrive on Earth within 24 hours after SFs. The five overlapped SWEs are kept in the analyses, otherwise the sample of SEPs will be too few. Even so, such overlapping will not alter the final conclusions of this paper. However, such overlapping could enhance the disturbance of the magnetosphere and ionosphere[3, 4, 30]. Accordingly, the associated compound effects of SWEs may partially explain why the SEPs affected flights have the most serious delays.

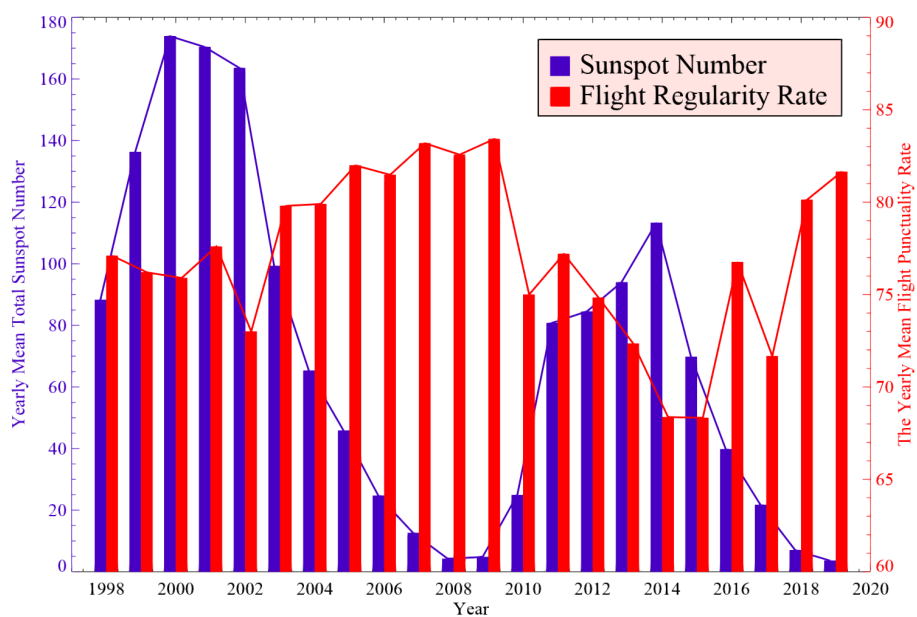

Fig.4. The yearly flight regularity rate of China(red) vs. the yearly mean total sunspot number (blue) from 1998 to 2019.

Although the contingencies that leading to flight delays have been smeared out to a considerable extent by the random distributions of SWEs and the usage of large amounts of flight data, we still intend to expend the research samples since longer research samples can also be conducive to reduce the impacts of various contingencies. Beside the 5-years' intact flight records, we will also examine the 22-years' national flight regularity rate data from 1998 to 2019. By analyzing the sunspot number and flight regularity rate data through entire two solar cycles (22 years), it is revealed that the yearly mean flight regularity rate was negatively correlated with the yearly mean total sunspot number. Sunspot number is an excellent indicator of solar activities, so a higher (lower) sunspot number means more (less) solar activities, and correspondingly, more (less) SWEs[4]. As shown in Fig.4, one can find that if there are more SWEs occur in a year, the flight regularity rate will be found to be lower. While the flight regularity rate tends to be higher if there are less SWEs. These results not only suggest that the long-term flight regularity rate could be modulated by SWEs but also confirm our previous results. Such consistencies also indicate that the methods we used to eliminate the internal periodicities and contingencies of flight delays are statistically valid, and the 'real' effects of SWEs on flight delays are successfully revealed.

When exploring the internal relationships between SWEs and flight delays, we have investigated many parameters related to the geomagnetic field and the ionosphere. Interestingly, although the intensity of geomagnetic storm is measured by Dst index, it is found that flight delays are better correlated with the dDst but not the Dst index directly as supposed. Other geomagnetic indices (SYM-H, AE, and Kp) have also been checked and none of them reveal a clearer correlation than dDst. Anyway, geomagnetic storm is a period of rapid magnetic field variations[28], and the results presented in Fig.3(a) illustrate a scenario that drastically increased flight delays tend to occur when the dDst becomes larger (especially dDst>20nT/hour). It also provides us a possible inference that more serious flight delays might be found in storm main phase than in storm recover phase since the dDst in storm recover phase is usually smaller than those in storm main phase during a geomagnetic storm. Geomagnetic storm-induced ionospheric disturbances have also been widely investigated by many researchers[3, 4, 8, 9, 26, 30]. The relationship between SWEs and the related magnetospheric-ionospheric disturbances has been demonstrated to be very complicated, and there is no one-to-one relationship between dDst, ΔfoF2 and ROTI. The foF2 data and TEC data are not easy to understand plainly since they are highly variable and depend on the time of the day, the season and the region. However, ΔfoF2 and ROTI are effective parameters that could be used to simply investigate the ionospheric disturbances. ΔfoF2>15% is usually considered as one of the typical indicators of obvious ionosphere disturbance (or even ionosphere storm) which would significantly interrupt the reliability and stability of radio communications[26]. While ROTI>0.2 TECU/min is often regarded as the existence of (apparent) ionospheric irregularities which would degrade or even interrupt the communication and navigation systems[29]. Although ionospheric disturbances are complicated to quantify, the obvious increases in flight delays shown in Fig.3(b) and Fig.3(c) are just in phase with the ΔfoF2 and ROTI when the disturbances above these two thresholds. Such consistent behaviors between flight delays and magnetospheric-ionospheric disturbances indicate that SWEs related negative effects on communication and navigation would increase the flight delay time and delay rate.

In fact, the impacts of SEWs on flight delays will be shown in many more aspects. For example, the flight departure delays during SWEs have also been analyzed by us, and the results are similar to the arrival delays shown here. SFs related flight delays reveal obvious latitude dependence and correlate well with the highest frequency affected by absorption[31]. SEPs will also affect the flight time on certain routes but the reason is not because of the safety concerns as usually assumed, e.g., the SEPs associated ionizing radiation or single-event error. These phenomena systematically imply that the influences of SWEs on aviation could be in many more ways that we do not understand. These detailed results are beyond the scope of this paper and will be presented in our subsequent studies[32, 33].

To sum up, flight delays are a major concern in civil aviation, because the delays would not only increase the airlines' additional economic costs but also reduce passengers' satisfaction. In order to improve the flight delay predications, it is very necessary and important to analyze the factors affecting flight delays. However, no one realized that flight delays would be systematically modulated by SWEs other than the SWEs associated flight safety issues. For the first time, our presented results quantity the delay effects and reveal the fact that compared with quiet periods, the flight delay time and delay rate during SWEs are significantly increased by 81.34% and 21.45% respectively. Further analyses suggest that the SWEs resulted magnetospheric-ionospheric disturbances could influence the aviation communication and navigation, which in turn leads to the increased the flight delays. These results expand the research field of traditional space weather and also bring us brand new thoughts to cope with flight delays.

## Acknowledgements

This work is jointly supported by the National Natural Science Foundation of China (41731067 and 42174199), Guangdong Basic and Applied Basic Research Foundation (2021A1515012581), and the Shenzhen Technology Project (GXWD20201230155427003-20200804210238001). We thank the International Service of Geomagnetic Indices, the Madrigal database and the Meridian Project Data Center for providing the Dst, TEC and foF2 data.

**Appendix**

The list of the SWEs is collected from below:

SFs:
https://www.ngdc.noaa.gov/stp/space-weather/solar-data/solar-features/solar-flares/x-rays/goes/xrs/

CMEs:
https://izw1.caltech.edu/ACE/ASC/DATA/level3/icmetable2.htm

SEPs:
https://umbra.nascom.nasa.gov/SEP/

All the selected SWEs are listed in Table A.

Table A. Lists of all SWEs

| SWE No. | Event Type | Start Date (YearMonthDay) | Start Time (HourMinute UT) |
|---|---|---|---|
| 1 | CME | 20150107 | 0700 |
| 2 | SF | 20150113 | 0413 |
| 3 | SF | 20150114 | 1230 |
| 4 | SF | 20150122 | 0443 |
| 5 | SF | 20150126 | 1646 |
| 6 | SF | 20150128 | 0421 |
| 7 | SF | 20150204 | 0208 |
| 8 | SF | 20150209 | 2219 |
| 9 | SF | 20150302 | 0631 |
| 10 | SF | 20150305 | 1706 |
| 11 | SF | 20150307 | 2159 |
| 12 | SF | 20150309 | 1422 |
| 13 | SF | 20150315 | 2242 |
| 14 | CME | 20150322 | 0200 |
| 15 | CME | 20150331 | 1800 |
| 16 | SF | 20150408 | 1437 |
| 17 | SF | 20150421 | 0708 |
| 18 | SF | 20150423 | 0918 |
| 19 | SF | 20150505 | 0942 |
| 20 | CME | 20150518 | 2000 |
| 21 | SF | 20150611 | 0849 |
| 22 | SF | 20150613 | 0720 |
| 23 | SF | 20150618 | 0033 |
| 24 | SEP | 20150618 | 1135 |
| 25 | SF | 20150620 | 0628 |
| 26 | SEP | 20150621 | 2135 |
| 27 | SF | 20150625 | 0802 |
| 28 | SEP | 20150626 | 0350 |
| 29 | SF | 20150703 | 1247 |
| 30 | SF | 20150706 | 0824 |
| 31 | CME | 20150713 | 0600 |
| 32 | CME | 20150807 | 1600 |
| 33 | CME | 20150815 | 2100 |
| 34 | SF | 20150821 | 0156 |
| 35 | SF | 20150824 | 0726 |
| 36 | SF | 20150828 | 1304 |
| 37 | SF | 20150830 | 0201 |
| 38 | CME | 20150908 | 0000 |
| 39 | CME | 20150913 | 0700 |
| 40 | SF | 20150917 | 0804 |
| 41 | SF | 20150920 | 0455 |
| 42 | SF | 20150927 | 1020 |
| 43 | SF | 20151001 | 1303 |
| 44 | SF | 20151004 | 0234 |
| 45 | SF | 20151015 | 22:45 |
| 46 | SF | 20151017 | 2009 |
| 47 | CME | 20151025 | 1400 |
| 48 | SEP | 20151029 | 0550 |
| 49 | SF | 20151031 | 1748 |
| 50 | SF | 20151104 | 0320 |
| 51 | CME | 20151107 | 0600 |
| 52 | SF | 20151224 | 0149 |
| 53 | SF | 20151228 | 1120 |
| 54 | CME | 20151231 | 1700 |
| 55 | SEP | 20160102 | 0430 |
| 56 | CME | 20160119 | 1000 |
| 57 | CME | 20160124 | 1800 |
| 58 | SF | 20160212 | 1036 |
| 59 | SF | 20160213 | 1516 |
| 60 | SF | 20160214 | 1918 |
| 61 | CME | 20160305 | 1900 |
| 62 | CME | 20160320 | 1400 |
| 63 | CME | 20160414 | 0900 |
| 64 | CME | 20160417 | 0300 |
| 65 | SF | 20160723 | 0146 |
| 66 | SF | 20160724 | 0609 |
| 67 | CME | 20160802 | 1400 |
| 68 | SF | 20160807 | 1437 |
| 69 | CME | 20161013 | 0600 |
| 70 | CME | 20161104 | 1800 |
| 71 | CME | 20161110 | 0000 |
| 72 | SF | 20161129 | 1719 |
| 73 | SF | 20170401 | 2135 |
| 74 | CME | 20170409 | 0000 |
| 75 | CME | 20170414 | 0000 |
| 76 | CME | 20170527 | 2200 |
| 77 | SF | 20170703 | 1537 |
| 78 | SF | 20170709 | 0309 |
| 79 | SF | 20170714 | 0135 |
| 80 | SEP | 20170714 | 0900 |
| 81 | CME | 20170822 | 0400 |
| 82 | SF | 20170904 | 0536 |
| 83 | SEP | 20170905 | 0040 |
| 84 | SF | 20170910 | 1535 |

| 85 | SEP | 20170910 | 1645 |
| 86 | SF | 20171020 | 2310 |
| 87 | CME | 20171225 | 0000 |
| 88 | CME | 20180309 | 2200 |
| 89 | CME | 20180513 | 0600 |
| 90 | CME | 20180606 | 1100 |
| 91 | CME | 20180625 | 0800 |
| 92 | CME | 2018030 | 2000 |
| 93 | CME | 20180710 | 1200 |
| 94 | CME | 20180825 | 1200 |
| 95 | CME | 20180923 | 0400 |
| 96 | CME | 20190507 | 2200 |
| 97 | CME | 20190511 | 0600 |
| 98 | CME | 20190514 | 0600 |
| 99 | CME | 20190516 | 2300 |
| 100 | CME | 20190527 | 0400 |
| 101 | CME | 20191029 | 2000 |
| 102 | CME | 20191102 | 2100 |
| 103 | CME | 20191111 | 1000 |

As seen from Fig.A, we choose 24-hour duration to define SWEs affected flights and quiet time flights to avoid the daily periodicity of the flight delays. Some flights data are not used if they are not in an entire day (the green area). It should be noted that the overlapping SWEs (Case B) are not counted in our analyses except the SEPs. In addition, any canceled flight is not used either. Moreover, unlike the SFs and SEPs that propagate very fast, it will take ~30 minutes for a CME from being observed (at L1 point) to the magnetosphere, so an additional 30 minutes' compensation is applied to all CMEs. Finally, all the time are converted to the local time (UTC+8) in data processing.

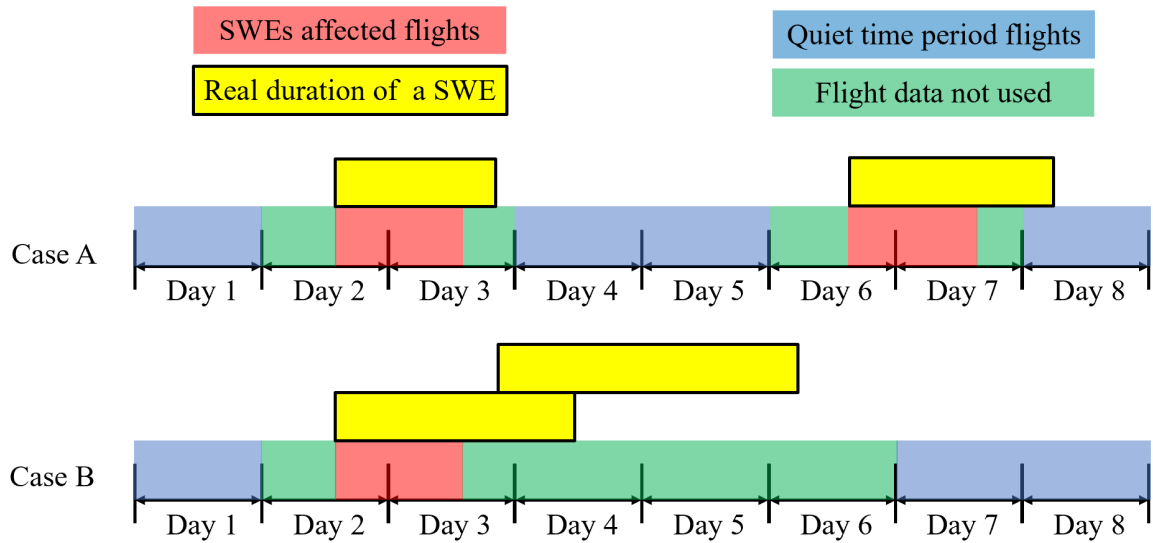

Fig.A. The definition of SWEs affected flights and quiet time flights.

The Dst index is obtained from the International Service of Geomagnetic Indices:

<div style="text-align:center">http://isgi.unistra.fr/data_download.php</div>

The TEC data is obtained from the Madrigal database:
<div style="text-align:center">http://cedar.openmadrigal.org/index.html</div>

The foF2 data is obtained from the Meridian Project Data Center:
<div style="text-align:center">https://data.meridianproject.ac.cn/</div>

  The TEC and foF2 data depend on the location and time. The location of the arrival airport and the arrival time are chosen to match each flight record. The TEC data is calculated by an automated software package to process GPS data based on a network of worldwide GPS receivers, and the vertical TEC data are adopted here. The space resolution of TEC data is 1° (latitude) × 1° (longitude) and the time resolution is 5 minutes. Since the TEC data usually have data gaps, we averaged the TEC data over a square 5°(latitude)×5°(longitude) region for each airport.

  The time resolution of the foF2 is 15 minutes. The regional foF2 data also have many data gaps as the TEC data, so we have to use the data from the other nearest stations to match each airport if the local data in unavailable. The data from Mohe station (E122°22′12.00″, N53°30′0.00″) is used to match PEK, Wuhan station (E114°36′36.00″, N30°31′48.00″) is used to match SHA and PVG, Hainan station (E109°7′58.80″, N19°31′33.60″) is used to match SZX and CAN.